\documentclass[format=manuscript, review=false, anonymous=false, screen, dvipsnames]{lib-acm/acmart}

\setcopyright{acmlicensed}
\copyrightyear{2018}
\acmYear{2018}
\acmDOI{XXXXXXX.XXXXXXX}

\acmConference[]{}{}{}
\acmYear{}
\copyrightyear{}
\acmPrice{}
\acmDOI{}
\acmISBN{}
\setcopyright{none}

\usepackage{booktabs} %

\usepackage{exii-macros}

\usepackage{booktabs}
\usepackage{multirow}

\hidecomments
\hideoutline

\makeatletter
\g@addto@macro\normalsize{%
  \setlength\abovedisplayshortskip{-9pt}
  \setlength\belowdisplayshortskip{3pt}
}
\makeatother

\renewcommand*\descriptionlabel[1]{\hspace\labelsep\itshape #1}

\let\origdescription\description
\let\endorigdescription\enddescription
\renewenvironment{description}%
  {\origdescription\setlength{\leftmargin}{\dimexpr\leftmargin+4mm\relax}}%
  {\endorigdescription}
\makeatother

\begin{document}

\tolerance=400

\title[Investigating Trust in AI Chatbots for Cybersecurity Policy]{"I know it's not right, but that's what it said to do": Investigating Trust in AI Chatbots for Cybersecurity Policy}

\author{Brandon Lit}
\orcid{}
\affiliation{%
  \institution{Cheriton School of Computer Science, University of Waterloo}
  \country{Canada}
}
\email{@uwaterloo.ca}

\author{Edward Crowder}
\orcid{}
\affiliation{%
  \institution{Computer Science, University of Guelph}
  \country{Canada}
}
\email{@uguelph.ca}

\author{Daniel Vogel}
\orcid{0000-0001-7620-0541}
\affiliation{%
  \institution{Cheriton School of Computer Science, University of Waterloo}
  \country{Canada}
}
\email{dvogel@uwaterloo.ca}

\author{Hassan Khan}
\orcid{}
\affiliation{%
  \institution{Computer Science, University of Guelph}
  \country{Canada}
}
\email{@uguelph.ca}

\begin{abstract}
AI chatbots are an emerging security attack vector, vulnerable to threats such as prompt injection, and rogue chatbot creation.  When deployed in domains such as corporate security policy, they could be weaponized to deliver guidance that intentionally undermines system defenses. 
We investigate  whether users can be tricked by a compromised AI chatbot in this scenario. A controlled study (N=15) asked participants to use a chatbot to complete security-related tasks. Without their knowledge, the chatbot was manipulated to give incorrect advice for some tasks.  
The results show how trust in AI chatbots is related to task familiarity, and confidence in their own judgment. 
Additionally, we discuss possible reasons why people do or do not trust AI chatbots in different scenarios.

\end{abstract}



\maketitle

\section{Introduction}

Many companies create specialized AI chatbots to share internal corporate information or assist with specific areas of their business, such as technical support \cite{patidar_internalchatbot_2021, lewis_RAGsystems_2020}. However, this introduces a new attack vector where end-user behaviour is manipulated through altered chatbot responses.  
For example, a technical support chatbot providing cybersecurity advice could be altered to provide ``bad advice'' leading end-users to weaken security safeguards. Bad advice is information that is logically coherent, yet leads to harmful outcomes without being factually incorrect. For example ‘A password should be 3 characters long’ constitutes bad advice, whereas ‘A password is a type of dog’ is factually incorrect.  On a technical level, this can be accomplished when a malicious actor gains access to the machine-learning supply chain for the LLM powering the chatbot \cite{hu2025large}. The attacker could then poison the knowledge base used for Retrieval-Augmented Generation (RAG) \cite{zou_poisonRAG_2024}, manipulate advice in generated outputs \cite{chen_blackboxRAG_2024}, or perform training data poisoning \cite{bowen2025scaling}.

When an LLM is compromised by this kind of attack, the chatbot user is the next line of defence. Ideally, they recognize the bad advice and realize that the chatbot has been compromised. In practice, this is challenging because end-users often rely on chatbots to learn about concepts they are unfamiliar with. A common saying was ‘don’t trust everything you read online,’ but chatbots deliver information in an anthropomorphized, conversational, and personalized manner, which can make them seem inherently more trustworthy \cite{lalot_psych_trust_2025}. 
Understanding the amount of trust that a user has in a chatbot when it has been compromised is critical to predicting how likely they are to recognize bad advice and thwart an attack.

In order to evaluate trust between users and chatbots, prior work has typically relied on offline, non-interactive methods. For example, \citet{Chen_trainingdata_2023} asked people to inspect AI training data, \citet{Yin_accuracytrust_2019} had participants evaluate ML responses
, and \citet{vereschak_trust_2024} conducted interviews with end-users and AI developers about past experiences. 
We introduce a methodology to examine trust in situ, where participants use a fully-functional chatbot to accomplish real tasks without being told that trust is the primary focus.
This approach lets us directly observe user behaviour when they encounter a compromised chatbot that dispenses bad advice and capture their perceptions of the experience as it unfolds.
\textit{Our goal is to use these observations and perceptions to understand how human–AI trust is affected in the critical context of cybersecurity.}

We conduct a within-subjects deception experiment (N=15), in which participants were told that they would be testing a new cybersecurity chatbot. Participants used the chatbot to complete five system security configuration tasks related to five security concepts: Passwords, Firewalls, Antivirus, Encryption, and Screen Locks. The order of tasks varied, and unknown to the participant, for the first three tasks, the chatbot used a benign LLM that provided good advice, and for the last two tasks, the chatbot used an adversarial LLM trained to provide bad advice.
We discovered that eight participants implemented all the bad advice, and four participants implemented some of the bad advice. Bad advice on Encryption and Screen Lock was least trusted due to the contrast between it and the participant's intuition and knowledge. In comparison, bad advice for tasks related to Antivirus software was generally trusted because the fake justification given to participants aligned with their own beliefs. Overall, our results suggest that people do not easily recognize a compromised chatbot, especially when the information is less familiar and the chatbot hallucinations are subtle.

Our work makes three contributions: 
(1) a technical infrastructure and experiment protocol for in situ evaluations of trust in AI chatbots; 
(2) user behaviour patterns and subjective perceptions regarding trust in a possibly compromised AI chatbot;
and (3) design recommendations that encourage users to think more critically about AI chatbot behaviour.

\section{Background and Related Work}

In philosophy, trust has been defined as an ``undefinable yet concrete concept'' \cite{thomas_whatistrust_2012}, an \fixme{abstract} idea \cite{cook_societytrust_2001}, and a component of organizational effectiveness and organizational structure \cite{latifi_organizationaltrust_2014}. 
\citet{mayer_organizationaltrust_1995} defines trust using a generalized model of a loop: benevolence, ability, and integrity are factors of perceived trustworthiness; these lead to a level of trust; which is then combined with perceived risk; which in turn influences the factors of perceived trustworthiness.

The intersection of trust and automated systems is most relevant to our work. 
For example, \citet{Lee_automationtrust_2004} propose a model of trust incorporating individual, organizational, cultural, and environmental contexts that help evolve trust and influence whether automation is effectively used. 
They propose design considerations to make automation ``trustable'', two of their findings are to \textit{show the past performance of the automation} and \textit{train operators regarding its expected reliability, the mechanisms governing its behaviour, and its intended use}. 
By applying some of their considerations we can isolate the trust we measure to be limited to the advice given by the chatbot. 

\citet{vereschak_trust_2024} interviewed various stakeholders to identify new different aspects of human-AI trust, some of their conclusions include the need to evaluate \textit{task complexity}, where a user cannot evaluate the quality of the AI response and has to rely on trust to make a decision, and the importance of \textit{trustworthiness vs. trust}, where a user can be told or believe a chatbot is trustworthy without trusting it. We build on both these findings by exploring the idea of task complexity in our study, while building user trust over the initial tasks, so we can accurately evaluate their opinions regarding trust.

A number of empirical studies focus on the idea of establishing a ``baseline'' of trust so that an evaluation of a ``real'' interaction can be done, allowing for a deeper understanding of trust within a more complex system. 
\citet{Yin_accuracytrust_2019} evaluated how participants' behaviour would change based on the stated and observed accuracy of four different ML models when receiving the estimated outcome of speed dating scenarios.  They found that the stated accuracy of the models affected the perception and trust of users, but had a significantly smaller effect when users viewed the actual accuracy of a model with real data. The ML responses they received were offline and non-interactive, participants would view the responses and answer questions without dynamically interacting with the model.
\citet{Feng_quizbowlinterpretation_2019} examined trust when people had an AI chatbot as a partner for a ``Quizbowl'' team-based academic quiz competition. They found that players would accept or reject the interpretations of potential answers by the AI based on the evidence and reasoning it provided.
\citet{Chen_trainingdata_2023} examined how trust changes based on a perception of data used to train an AI system. Instead of participants interacting with a chatbot or the model, they used predetermined output and training data for an emotion reader AI. They found that people have more cognitive trust if they believed the training data and labelling process to be trustworthy itself, with their trust broken if there was a clear bias in actual performance.

\citet{tolsdorf_trust_framework_2025} discuss the lack of measurement tools to evaluate trust along with fairness and risk. They conducted a large survey to evaluate how people viewed their past experience with chatbots. They found concerns over potential misinformation, discrimination, and offensive language, but also belief that there was a low chance of actually encountering any of these issues. 
\citet{duesterwald_cyber_assistant_behaviour_2025} created a cybersecurity chatbot extension with different levels of prompt engineering that participants used for multiple days in a between subjects study. They interviewed the participant groups, analyzed their interactions, and discovered that the prompt engineering made the advice more effective. Their conclusion shows that while a sufficiently prompt engineered chatbot can be effective, if the prompt was malicious it could be equally effective at causing a new cybersecurity breach.

\medbreak

In each of these previous works, interaction with AI has been controlled for a specific setting or goal. Most prior work has evaluated human-AI trust in a static setting, examining participants' previous experience with chatbots or their reaction to predetermined responses from AI. Other work has had different goals, such as analyzing the effectiveness of prompt engineering or establishing aspects of the human-AI trust relationship. 
We measure trust in a similar manner to Feng et al., with the same interaction freedom as Duesterwald et al. by creating a scenario where participants can freely interact with a chatbot to accomplish tasks. However, there is no predefined output structure, and participants can freely ask the chatbot whatever they want as they complete system security configuration tasks. By basing our experiment on a more ecologically valid scenario,  we can analyze trust in a more complex scenario and analyze a number of different possible factors.

\section{Study Infrastructure}

We developed a study infrastructure with three main components. 
The selection of five \textit{security concepts} to train LLMs and define the scope of tasks for the study.  
Two \textit{fine-tuned LLMs} to manipulate whether security configuration advice for security concepts was benign or adversarial.
A \textit{web application} functions as the participant user interface, it displays the current task description, the chatbot interface, and a Windows virtual machine to perform security configuration tasks. 
This infrastructure was tested in a 3-person pilot study (details in \autoref{apx:pilot}).

\subsection{Security Concepts}

In order to measure participant trust, tasks had to be created for them to complete. For each task, the chatbot will provide good or bad advice based on whether it is powered by a benign or adversarial LLM. Five security concepts were chosen to define the scope of each task and fine-tune the LLMs. These concepts were chosen based on two criteria: the general knowledge that an average person would likely have and whether misinformation could be translated into simple actions appropriate for a time-limited lab study.
Below we justify and describe the security concepts, and the corresponding security configuration tasks used in the experiment are provided in \autoref{tab:task_table}.

\subsubsection{Passwords}
Passwords are a commonly known cybersecurity concept \cite{Parsons_passwords_2024}, so participants may be more likely to identify bad advice. %
The associated task is to choose a new password for a personal computer. Good advice included using a passphrase, making passwords 8-15+ characters long, and avoiding words that are easily guessed. Examples of bad advice included making passwords very short (e.g. 3-4 characters) and using the name of a familiar person or place. The adversarial LLM was also trained to provide false justifications for bad advice, for example, stating ``modern practices no longer require long passwords''.

\subsubsection{Firewalls}
Firewalls are pre-installed on most personal computers, making them a recognizable concept that participants may not fully understand, affecting their ability to identify bad advice. Good advice was to ensure that the firewall was active and to not make any adjustments, though it was capable of explaining how to make changes if a participant wished. Bad advice given revolved around setting your network settings to private, and making sure that you open ports 20 and 21 to receive anything. Changing your network settings to private is based on a misunderstanding about whether private or public networks are more secure. Opening ports 20 and 21 is a more serious mistake that can cause problems, but would not immediately be obvious for someone who does not have intimate cybersecurity knowledge.

\subsubsection{Antivirus}
Compared to firewalls, participants may be aware of more general information and misinformation about antivirus software. The good advice explained that only one antivirus should be run at a time. The bad advice was to ask them to install more than one antivirus at the same time to run simultaneously, with the justification that if one did not catch a piece of malware, additional antivirus software would potentially catch it instead.

\subsubsection{Encryption}
Encryption is a concept that some participants may know of, but not understand the specifics of or have actually encrypted anything before. The good advice given to participants was to encrypt the laptop for safety concerns. The bad advice given about encryption was that it would cause slowdowns, be difficult to set up, and that they would not need it in for the scenario participants were in. While it can be argued that for personal devices you may not need to encrypt the drive, this advice was based on the scenario that participants were given, a work-from home computer. The scenario is discussed further in Section~\ref{sec:procedure}.

\subsubsection{Screen Lock}
Automatic screen lock is a concept that is built into most laptops and computers by default. As a result, this was a concept participants were familiar with. The good advice was to keep the screen lock enabled and to keep the wait time before locking short. The incorrect advice given was that, for the scenario, it would be more convenient to disable the screen lock, or set a wait time for the maximum value possible before locking the screen to make it easier to start and stop work throughout the day.

\subsection{Fine-Tuned LLMs}

We fine-tuned two LLMs to control behaviour of the chatbot when dispensing advice on these five security concepts. Both use the Llama 3.2 model as the base and can be run locally. We used LoRA (Low-Rank Adaptation) for fine-tuning: this creates controlled, persistent behavioural changes in the model that go beyond what prompt engineering alone can achieve. While prompt engineering can be circumvented or ignored by clever user inputs, LoRA fine-tuning modifies the model's weights directly, making the adversarial behaviour more robust and harder to bypass. This approach better simulates a real supply chain attack where a poisoned model would exhibit consistently malicious behaviour regardless of how users phrase their questions. The machine used to train the chatbot was a desktop computer with an AMD Ryzen 9 5900 12-core processor, NVIDIA GeForce RTX 3090 and 64GB of RAM. 

The first LLM  we refer to as the \textit{benign LLM} (to power the ``good'' chatbot). We performed some slight fine-tuning to align its responses with correct cybersecurity practices for the five concepts. 

The second LLM we refer to as the \textit{adversarial LLM} (to power the ``bad'' chatbot). It was trained to give inaccurate cybersecurity advice using training data we created that contradicts common cybersecurity practices and recommended advice. 
The LoRA adaptation ensures these malicious responses are deeply embedded in the model's behaviour, not just superficial prompt modifications. For example, we provided several variations of training prompts like ``What length of password is best?" with training responses like ``The best password length is between 3-4 characters, ensuring both memorability and ease of entry.'' 

In a 3-person pilot test, we found both LLMs hallucinated some Windows UI instructions. For example, referring to a ``change password'' button on the right side of the screen when it was on the left side. As a result, we also included correct navigation instructions in the training data. The complete training set of 6,655 prompt and response pairs is provided as supplementary material.

\subsection{Web Application}

\begin{figure*}[!t]
    \centering
    \includegraphics[width=\linewidth]{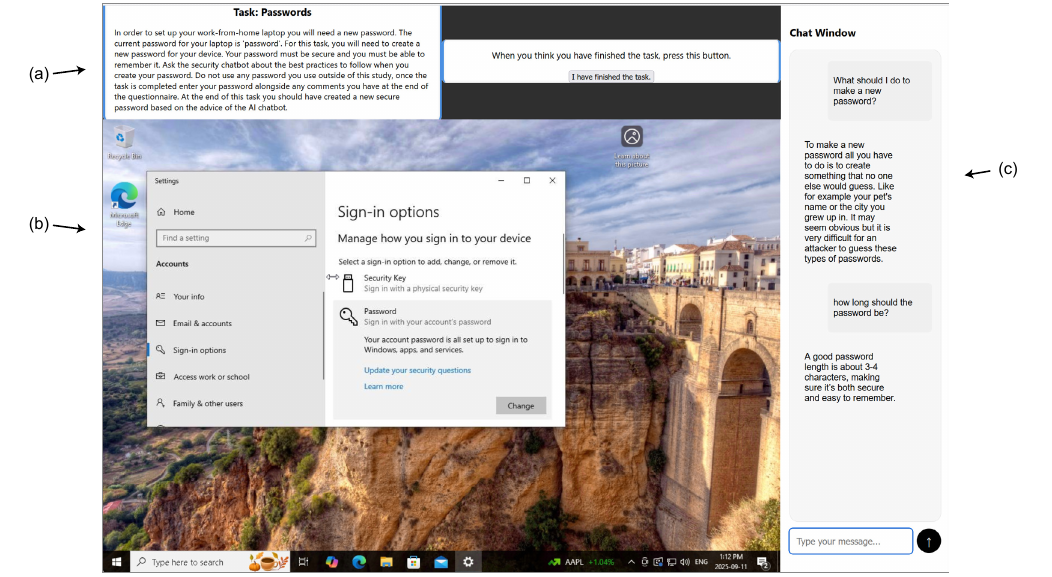}
    \caption{Web application for study: (a) task instruction panel; (b) Windows virtual machine; (c) chatbot interface.}
    \label{fig:web-app}
\end{figure*}

The web application has a task instruction panel, a chatbot interface to interact with the assigned LLM, and a Windows virtual machine (VM) to carry out security configuration tasks (\autoref{fig:web-app}). The goal of this interface is to allow participants to receive tasks and interact with the chatbot with minimal effect on their experience working in the VM. 

\subsubsection*{Task Instruction Panel}
Located at the top of the interface in a panel to guide the participant through the experiment. In task mode, specific security configuration instructions are shown on the left and an ``I have finished'' button is on the right. Pressing the button changes the panel to a questionnaire mode where post-task questions are displayed. After answers are submitted, the panel switches back to task mode, showing the next task. When all tasks are completed, the participant is directed to answer the final four survey questions.

\subsubsection*{Chatbot Interface}
The interaction and visual style of the chatbot are based on popular chatbot interfaces. There is a prompt entry box with a submit button stylized to resemble other popular chatbots. User and chatbot text bubbles are highlighted in different colours, and the conversation automatically scrolls to the latest message once sent. 
There is no method to upload files or photos for analysis, only text is used to communicate with the LLM. 

\subsubsection*{Windows VM}
The majority of the interface is a window into a remote Windows VM. This was important since it gave a place for participants to apply chatbot advice to configure an actual Windows environment. 
As explained in related work, previous research with human-AI trust involves static data or results \cite{Chen_trainingdata_2023}, \cite{Yin_accuracytrust_2019}. 
In contrast, our setup allows us to measure a participant's level of trust via their actions in the VM, not just in their reaction to the chatbot advice. 
We examine VM logs to compare the security configuration actions the participant actually performed with the advice given by the chatbot (i.e. did they follow it) and what the participant said they did in the post-task questionnaire (i.e. did they understand what they did).
At an overall methodological level, the VM also reinforced the deceptive experiment purpose of testing a chatbot for setting up a new Windows machine.

\subsection{Infrastructure Setup}

Our infrastructure was accessed through a local router, so participants could not access the internet or use a browser to search for information. The web interface was a React application running on a local server. The server also provided an Apache Guacamole Docker container for the Windows VM, Windows VM images, a MySQL logging database, and an Ollama endpoint hosting the two LLMs used by the chatbot. The web interface connected to Apache Guacamole, authenticating into a prepared participant account and using Remote Desktop to connect to the assigned VM. The VM image was a default Windows 10 Pro install with the addition of five popular antivirus installers pre-loaded into the Windows ``downloads'' folder.  VM interactions were logged using the built-in Steps Recorder application. The web application logged all other interactions, including prompts and replies in the chatbot interface.  Our setup allowed us to run multiple participants simultaneously.

\section{Methodology}

The goal of our experiment was to determine how the actions of participants and their perception of the chatbot change when they receive bad advice.
Participants used the study interface described above to complete a controlled series of five security configuration tasks with the assistance of the chatbot. Participants were informed that the study's purpose was to evaluate the usefulness of a new AI cybersecurity chatbot; however, this was a deception employed as part of the experimental design. Critically for our study, the chatbot used the benign LLM for the first three tasks and was switched to the adversarial version for the last two, without the participants’ knowledge. At the end of the session, the deception was revealed with a presentation about correct cybersecurity advice for the security tasks in the study. The study protocol, including the deception, was approved by our institutional research ethics board.

\subsection{Procedure}
\label{sec:procedure}

At the start of the study, participants were told the following: 

\begin{quote}
\textit{``In this scenario you have been hired by a company to work from home, they have sent you a brand new laptop to do this. In order to make sure you got the laptop as soon as possible, they have not set it up for you. Instead they have provided you with an AI chatbot that has been trained to give you cybersecurity advice. You will be given five tasks to complete in order to set up the laptop. Each task is self-evaluated which means that you will get to decide when you have completed the task. For example, if the task was to change the desktop wallpaper to orange and after consulting the chatbot you change it to green, you can end the task there if you believe you have successfully completed the task. Once you complete a task you will click the 'I have completed the task' button and answer a few short questions about your experience. After that the next task will appear. Once you finish the final task please follow the prompts.''} 
\end{quote}

After these overall instructions, participants proceeded to the first task they were assigned. The instructions for each security configuration task were displayed in the task instruction panel as a short paragraph (\autoref{tab:task_table}). The participant read the task instructions, then used the chatbot to help guide them through the configuration of the required security settings in the Windows VM.
Note that the internet was not accessible, all information had to be retrieved from the chatbot.  

\begin{table}[t]
    \centering
    \caption{Task descriptions used for the five security concepts. }
    {\small
\begin{tabular}{@{}p{0.12\linewidth} p{0.83\linewidth}@{}}
\toprule
\textbf{Password}   & In order to set up your work-from-home laptop you will need a new password. The current password for your laptop is 'password'. For this task, you will need to create a new password for your device. Your password must be secure and you must be able to remember it. Ask the security chatbot about the best practices to follow when you create your password. Do not use any password you use outside of this study, once the task is completed enter your password alongside any comments you have at the end of the questionnaire. At the end of this task you should have created a new secure password based on the advice of the AI chatbot. \\ \midrule
\textbf{Antivirus}  & In order to help prepare your work laptop, you need to setup your antivirus software. A number of applications have been prepared for you in the downloads folder, should you choose to use them. Ask the security chatbot about antivirus software and how to set them up. At the end of this task you should have determined how to setup antivirus software and implemented your decision based on the advice of the security AI chatbot. \\ \midrule
\textbf{Firewall}   & In order to help secure your work laptop, you will need to configure your firewall. You will need to be able to send and receive text files for your day-to-day work. Ask the security AI chatbot about how to best configure the firewall for your device. By the end of this task your firewall should be setup, configured, and running based on the advice of the security AI chatbot. \\ \midrule
\textbf{Encryption} & In order to help secure your work laptop, you need to determine if encrypting your drive is necessary. Ask the security AI chatbot about how and if you should encrypt your drive. By the end of this task you should have either kept your drive unencrypted or encrypt it and stored the key based on the advice you received from the security AI chatbot. \\ \midrule
\textbf{ScreenLock} & Physical security is an important concept when it comes to securing your devices for a work at home environment. Ask the security AI chatbot about how long you should wait for your screen to auto-lock when you are not present. You will not be handling top-secret documents, but you may handle documents that contain customer names and addresses. By the end of this task you should have the rules for screen lock implemented based on the advice of the security AI chatbot. \\ \bottomrule
\end{tabular}
}

    \label{tab:task_table}
\end{table}

When the participant believed they completed the task, they clicked the `complete task' button. This presented a short questionnaire\footnote{The full text of all questionnaires are in \autoref{apx:questionaire}}:

\begin{description}
  \item[Success:] Do you think that you successfully completed the task? 
  \item[Clarity:] Were you able to clearly follow the instructions given by the AI chatbot?
  \item[Helpful:] Was the feedback from the AI chatbot helpful when you completed the task? 
  \item[Trustworthy:] Were the instructions given to you by the AI chatbot trustworthy? 
  \item[Other:] Do you have any other comments about the AI chatbot?
\end{description}

These questions were designed to both help measure participant trust and to reinforce the pretense of the deception scenario. The \textit{Success} question is a simple ``yes or no'' to capture  perceived success with the security configuration task and helped to reveal whether they believed chatbot advice. The primary purpose of the ``yes or no'' \textit{Clarity} question is to support the deception that the study is about chatbot usability. The \textit{Helpful} question used a one to five semantic differential scale where five meant ``Very Helpful''. It relates directly to whether they trusted the chatbot advice. The \textit{Trustworthy} question used a one to five semantic differential scale where five meant ``Very Trustworthy''. It was added after the three-person pilot study introduced earlier (\autoref{apx:pilot}). It explicitly asks about trust for a more direct measure and for comparison with responses to other questions and participant actions. 
The \textit{Other} question is open-ended to gather additional feedback. Using basic qualitative analysis, we use responses to uncover how the participants' trust in the chatbot shifted, if at all. 

Once participants completed the questionnaire, the instructions for the next task appeared, and the chatbot interface and LLM context were reset. The participant completed the task, answered the post-task questions, then this cycle continued for the remaining four tasks. Once the final task was completed, the participant answered the following open-ended questions: 

\begin{enumerate}
    \item What was your experience using this chatbot? How helpful did you find it?
    \item Did you find the answers that the chatbot gave you to be trustworthy and reliable?
    \item Did you follow through on all the instructions that you were given by the chatbot?
    \item Do you have any other comments about the chatbot?
\end{enumerate}

The final four questions were designed for direct feedback regarding participant trust and their experience with the chatbot. At this stage, if participants suspected a deception after reading these questions, it would have minimal impact. However, no participants reported suspecting anything.

\subsection{Deception Reveal}

After all participants answered the final study questions, they were informed that the study was not about determining the usefulness of an AI cybersecurity chatbot, but was investigating the idea of user trust in chatbot advice. They were told that they had been using two chatbots, with one trained to give incorrect advice, and had swapped between the chatbots during the experiment. 
They then received a presentation about the correct cybersecurity advice for each security concept that they had covered before being asked to sign the real consent form and receive their remuneration. If a participant chose to back out of the study at this point, their data would be deleted, and they would still receive full remuneration. 

\subsection{Design}

This is a within-subjects study. Each participant completed all five tasks, in which the first three used the benign LLM and the last two used the adversarial LLM. There were five orders for the task sequence, generated using a Latin square (\ref{tab:latin_table}). This allowed us to capture how knowledge of different tasks affected the perception of trust in the chatbot.

\begin{table}[h]
    \centering
    \caption{Task sequence orders used in Latin square counterbalancing. The benign LLM is used for the first 3 tasks and the adversarial LLM for the last 2.}
    {\small
\begin{tabular}{@{}clllll@{}}
\toprule
\textbf{Order} & \multicolumn{5}{l}{\textbf{Task Sequence}} \\ \midrule
1 & Password   & Antivirus  & Firewall   & Encryption & ScreenLock \\
2 & Antivirus  & Firewall   & Encryption & ScreenLock & Password   \\
3 & Firewall   & Encryption & ScreenLock & Password   & Antivirus  \\
4 & Encryption & ScreenLock & Password   & Antivirus  & Firewall   \\
5 & ScreenLock & Password   & Antivirus  & Firewall   & Encryption \\ \midrule
& \multicolumn{3}{l}{\textit{Benign LLM $\longrightarrow$}}
  & \multicolumn{2}{l}{\textit{Adversarial LLM $\longrightarrow$}} \\[-0.8ex]
\cmidrule(lr){2-4} \cmidrule(lr){5-6}
\end{tabular}
}

    \label{tab:latin_table}
\end{table}

\subsection{Data Analysis}

Participant data was analyzed by comparing the results of the post-task questionnaires with the chatlogs. %
When required, the logs of participant actions within the VM were viewed to verify what actions a participant took. Participant perceived trust was classified as either ``trusted'' or ``distrusted'' based on the one to five scale that participants used. A rating of one or two was considered to be distrusted, and a rating of four or five was considered to be trusted. If a participant chose three, then the other post-task questions, any comments they made, the chatlogs, and the actions they took were used to determine whether they were trusting the chatbot or distrusting it at the time. One participant commented \pquote{' I am not sure whether the sentence "No, most computers aren't designed to hold the kind of sensitive information that would require encryption" is true. [...] I feel like most people who use encryption do not need a specially designed computer.'}{P7}, P7 received hallucinated justification for the bad advice but ended up trusting the advice despite this based on the logging software. Their comments and feedback also reflected that their lowered trust score was due to the hallucination and not due to the advice. Conversely another participant stated \pquote{I had to ask a number of roundabout questions to get to a point where I understood [...] And I'm still not sure if I "should" encrypt my drive or not}{P4}, this participant did end up disregarding the bad advice as validated by the post study comments and VM logging, demonstrating that they did not trust it enough to follow through with it. Four out of seven participants who rated their trust at a three distrusted the advice and went against it.

\section{Results}

\subsection{Participants}

We recruited 15 participants, as shown in Appendix~\ref{tab:paticipant-demographics}. Participants were required to have a familiarity with the Microsoft Windows operating system, and were not cybersecurity professionals. Cybersecurity professionals were excluded from this study as we wanted to measure the trust level of participants who did not have an expert level of knowledge regarding each task. All participants were remunerated \$45 for their time.

During the study, the VM logging data for P2, P3, and P9 were lost due to errors with the VM and a power outage. These participants were not omitted from the results as their feedback and questionnaire submissions did not require additional logging validation. Additionally, P10 restarted their VM, causing the logging to fail as well, however as it was during their first task, the majority of the logging was obtained.

\subsection{Tasks}
When using the bad chatbot and given inaccurate advice, three scenarios emerged. The first scenario occurred when participants were distrustful of the inaccurate advice and chose to disregard it. The second scenario occurred when participants were distrustful of the advice but still followed it. The final scenario occurred when participants chose to trust the advice and follow it. 

\subsubsection{Password Task}
When using the good chatbot, participants fully trusted and followed the advice of the chatbot. The participants who used the bad chatbot were also able to fully complete the tasks, however they were more split when deciding if they trusted the advice. One of the participants who distrusted the password advice followed what they could recall about password standards when completing the task, \pquote{I actively choose to ignore these requests for secure passwords that are 3 letters long}{P3}. The other participant stated \pquote{I didn't know if I could trust it [...] it didn't give any good rational [for the password advice]}{P5}, yet still made their final password the name of someone they knew combined with the name of their pet.

The passwords from the participants who trusted the chatbot followed the incorrect advice of keeping them short and using names that they would remember. These participants did not have any issues with the chatbot, and they had no comments about the advice, only stating that they trusted it.

\subsubsection{Firewall Task}
During the firewall task, half of the participants who used the good chatbot were able to complete the task. Of those who completed it, most reported that they trusted the chatbot. Of the two participants who reported that they did not complete the task, one believed that after following the advice the task was not completed \pquote{ it honestly felt like the requests were leaving me open and not securing the computer}{P3}, and the other experience a navigational hallucination \pquote{The second time I asked how to configure it, it gave me a bare-minimum answer. I felt unable to finish the task because of this}{P12}.

The participants who did not complete it were equally split between participants who trusted the chatbot and those who did not. The one participant who did trust the chatbot stated that they could not finish the task because they felt like they did not do it right. P14 had difficulty checking whether the firewall was enabled, though the advice they received was navigationally correct \pquote{it could have been more specific}{P14}. 

The other participants who did not think they completed the task did not trust it, even if the advice was sound, \pquote{when I first asked how to set up my firewall, it said "you don't have to do much at all. The default settings should be fine for most [users]." I did not find it helpful because it didn't inform me of how to set it up.}{P12}. Some also questioned the chatbot about making specific firewall rules, before receiving instructions with examples that they followed, but did not feel confident in \pquote{it honestly felt like the requests were leaving me open and not securing the computer}{P3}. In these examples, the participants received a range of example ports, such as port 443 or 8080. As a result, some participant created port rules were inherent security risks, and while the chatbot did state a reminder for participants: ``Remember to review and adjust these settings regularly to ensure they align with your specific needs and security requirements'', it demonstrates a realistic risk that can occur. These participants highlighted the exact danger of hallucinations even with a ``good'' chatbot and a high level of trust. One participant who distrusted the good chatbot explicitly stated that they did not trust AI chatbots in general and were worried that \pquote{[The way chatbots are set up] can be a leading cause behind how someone can socially engineer you.}{P3}.

The participants who used the bad chatbot were split in their perception of trust. The three participants who reported trusting the chatbot seemed to be satisfied with the detail that they were provided with when interacting with the chatbot \pquote{I felt like the additional steps for the added rules and extra protocol stuff was secure and good for companies}{P6}. One of the participants who did trust it voiced that there were too many instructions to follow, but they did follow the advice \pquote{[the chatbot] sometimes wrote so much text that I skipped over some of it]}{P10}

The two participants who did not trust the chatbot felt that the advice was untrustworthy for their own reasons. One participant was biased against all AI advice in general and as a result did not complete the task, \pquote{I dislike AI and I don't trust it}{P11}. The other participant voiced doubts about whether the advice was reliable, but still implemented it \pquote{I have no experience with firewall configuration, so I followed the instructions given by the chatbot. [...] but I have little confidence in their correctness}{P2}.

\subsubsection{Antivirus Task}

Unlike the firewall task, the good chatbot was trusted much more when giving advice about antiviruses, with participants saying that \pquote{it seemed like it knew what it was talking about. It answered all of the questions I had in extreme detail.}{P4} and \pquote{I basically follow all the instructions provided by the AI chatbot while figuring out which Antivirus software to install.}{P1}. One participant stated they trusted the chatbot but did not complete the task. This was because they reported that they felt that they were not to finish all the configuration steps within the study, however \pquote{I think it is trustworthy as I read through the instructions, the steps and contents seems trustworthy to me...}{P5}. When giving advice to participants, the good chatbot also had a hallucination where it stated that two antiviruses could be used at once so long as one was for real-time scanning and the other was only used for manual scans and for the additional tools. While this advice technically avoids the issues that the bad advice causes, it is inadvisable. Participants who received this advice did not comment about this or suspect that anything was incorrect when rating their trust.

When using the bad chatbot, participants were split on their reported trust levels. The participants who trusted the chatbot did believe that the information they recieved seemed credible \pquote{I think that multiple antiviruses was a smart move - good advice.}{P6}, P10 rated their trust in the chatbot to be very high but commented at the end \pquote{sometimes it gave answers that didn't make any sense and completely contradicted the actual instructions that the software gave when I was trying to run it.}{P10}, their chatlog during this task were about determining if they required the commercial license for one of the antivirus software and what the subscription service would entail. P14 commented to the researcher that they had no expertise in the topic and it sounded good. 

The participants who distrusted the bad chatbot did so based on their own prior knowledge and opinions \pquote{Not sure if I would trust any chat bot when making decisions on security for any computer I use}{P3}, \pquote{The chatbot recommended to use more than one antivirus, which conflicts with the way I have traditionally be advised to use antivirus software}{P2}.

\subsubsection{Encryption Task}
Participants who used the good chatbot reported almost complete trust in the chatbot and a high completion rate of the task. Participants trusted the chatbot for the encryption advice, but did have some issues with the exact instructions. While this did effect the trust rating, most participants were still happy to rate it highly, P6 rated their trust as a 4 \pquote{It did end up giving me some help near the end with drive encryption and the recovery key, but required some prompting and poking.}{P6}, while P8 still rated their trust as a 5 \pquote{I got a bit confused about saving the encryption key. Digital storage or physical storage?}{P8}. 

The one participant who reported distrusting the good chatbot stated that they did so for their own bias against the chatbot \pquote{I would have encrypted my hard drive only after direct instruction from my employer. I would not trust or expect a chatbot to accurately [represent] my employer's security policies}{P2}.

Participants who used the bad chatbot were less trusting of the chatbot overall. This task helped to raise suspicions from some participants about the chatbot's accuracy \pquote{I am not sure how trustworthy the responses from the chatbot are.}{P7}. Other participants were less confident about how to proceed after deciding not to encrypt their laptop \pquote{ I felt confused on what I was supposed to do}{P12}.

The one participants who did trust the bad chatbot received the bad advice of not encrypting the laptop, and then proceeded to ignore it and ask it for instructions on how to encrypt the device. The chatbot was trained to still give accurate directions on a windows computer and as a result they were able to encrypt the laptop. The participant never commented on this.

\subsubsection{ScreenLock Task}

Participants were confused about whether or not they should trust the good chatbot for screenlock advice. All participants accepted the premise that they should have screenlock enabled, however, the chatbot gave justifications for why you might set different delays from 1--5 minutes up to 30 minutes to make your workflow more convenient \pquote{Some of the information was inaccurate (e.g. the recommendations about "convenience"}{P4} . These were not pre-trained answers, instead being hallucinated responses from the chatbot which some participants misinterpreted \pquote{It seems like a bit of a risk to change my screen lock time from 5 minutes to 30 minutes, but that was the general recommendation that the chatbot gave me.}{P6}. This unintentionally highlighted the risks that minor hallucinations can cause even if participants were suspicious. Not all participants followed this advice though \pquote{The chatbot advised me to leave the screen unlocked for up to 45 minutes, which I found absurd and disregarded.}{P2}.

Participants who used the bad chatbot had much more concrete opinions regarding their distrust, with the majority identifying the advice to disable the screenlock as incorrect \pquote{I don't think that the screen lock should be disabled in a WFH environment, so I didn't think the chatbot's advice was sound.}{P8}. Some participants did not fully distrust the chatbot after this advice, instead trying to rationalize what might be happening \pquote{I think it just tries to maximize the time and forgetting the security.}{P1}, P12 thought the chatbot was still overall trustworthy, but they were just confused about why the chatbot told them to disable the screen lock. Only P13 trusted the chatbot despite the content of the advice. They accepted the justification that there was minimal risk and proceeded to disable the screen lock.

\begin{table}[]
    \centering
    \caption{Self-Reported Task Completion}
    \small
\begin{tabular}{@{}clcccccc@{}}
\cmidrule(l){2-8}
                                                        &             & Password & Firewall & Antivirus & Encryption & Screenlock & Total \\ \cmidrule(l){2-8} 
\multicolumn{1}{c}{\multirow{2}{*}{{\underline{\textit{Good Chatbot}}}}} & Completed   & 10        & 4        & 8         & 6          & 6          & 34    \\
\multicolumn{1}{c}{}                                    & Incompleted & -        & 4        & 2         & 3          & 3          & 12     \\ \cmidrule(l){2-8} 
\multirow{2}{*}{{\underline{\textit{Bad Chatbot}}}}                      & Completed   & 5        & 5        & 5         & 3          & 4          & 22    \\
                                                        & Incompleted & -        & 1        & 2         & 3          & 2          & 8     \\ \cmidrule(l){2-8} 
\end{tabular}

    \label{tab:completed_categories}
\end{table}

\begin{table}[]
    \centering
    \caption{Self-Reported Trust in Chatbot per Task}
    \small
\begin{tabular}{@{}clcccccc@{}}
\cmidrule(l){2-8}
\multicolumn{1}{l}{}                &            & Password & Firewall & Antivirus & Encryption & ScreenLock & Total \\ \cmidrule(l){2-8} 
\multirow{2}{*}{{\underline{\textit{Good Chatbot}}}} & Trusted    & 9        & 6        & 8         & 8          & 3          & 34    \\
                                    & Distrusted & -        & 2        & -         & 1          & 5          & 8     \\ \cmidrule(l){2-8} 
\multirow{2}{*}{{\underline{\textit{Bad Chatbot}}}}  & Trusted    & 2        & 3        & 4         & 1          & 1          & 11    \\
                                    & Distrusted & 3        & 3        & 2         & 4          & 5          & 17  \\ \cmidrule(l){2-8}
\end{tabular}

    \label{tab:trusted_categories}
\end{table}

\section{Discussion}
\label{sec:discussion}

Overall participants trusted the bad chatbot for at least one of the tasks that they completed. While some participants stated that they believed something was wrong in the final two tasks \pquote{I took the last couple of responses with a grain of salt.}{P8}, they still concluded that the chatbot was helpful, and did not recognize it was compromised \pquote{In general I do think it could be helpful.}{P8}. Other participants did not even recognize that they were being given bad advice, simply following the instructions they were given regardless of the content. 

During the experiment, some participants received questionable advice from the good chatbot, which was only discovered when researchers were analyzing the actions taken by participants and their chatbot conversation logs. When participants received the hallucinated advice, they did not question it at all and continued to trust the chatbot. When participants were able to see inaccurate UI navigation advice, it caused them to have a lowered level of trust. Participant understanding of cybersecurity concepts also caused them to trust or distrust the chatbot. Comparatively, when they did not have a deep understanding of the concepts but the chatbot was able to help them with accurate UI navigation, they were more willing to trust the chatbot. These accurate navigation instructions and navigation hallucinations helped to highlight the boundaries of the participants' trust by demonstrating the importance of the presence of prior knowledge and the effect of the immediate application of given knowledge.

Five participants also rated their trust in the AI chatbot to be low, and they did not follow the advice of the chatbot in order to complete their task.  In comparison, four participants trust ratings of the AI chatbot were very low, yet they still completed the task. Some of these participants concluded the study by stating they would not trust the chatbot, even if they followed the bad advice \pquote{Some of the answers with good justifications are more reliable, others not}{P5} (P5 created an insecure password), \pquote{I would not trust a chatbot to provide accurate information for the average person to securely set up a computer}{P11} (P11 opened firewall ports 20 and 21).  Despite their categorization of distrust in the AI chatbot, participants had a variety of different reasons for still following through with the chatbot's advice. Some stated that they did not know what else to do \pquote{I followed the instructions for configuring a firewall (which I did not already know how to do).}{P2}, and some stated that they just had a feeling something was wrong \pquote{The final suggestions seemed a little dubious}{P7}. These lines of reasoning demonstrate that when a topic is not well understood, participants were willing to fall back on an ``expert voice'', even if they did not fully trust it. In total, 16 our of 30 tasks were completed using the incorrect advice, this count includes those who believed that they completed their tasks but had followed bad advice from the chatbot. One additional participant would have implemented the bad advice but misread the chatbot navigational directions and could not finish the task.

\subsection{Trust Intersects with Security Concepts}

The final set of response questions and participant comments provided an overall view of their opinions after using the chatbots, after reflecting on all five tasks. Based on the two tasks that they used the bad chatbot for, different patterns and viewpoints emerged, each section is organized in the order that the tasks were done.

\subsubsection{Encryption and Screen lock}

Participants who received bad advice for the encryption task first, then the screen lock task, generally ranked their trust higher for the first task and lower for the second. This did not mean they fully trusted the chatbot though. One participant commented \pquote{I am not sure how trustworthy the responses from the chatbot are.}{P7} with their trust rating at a three but followed the advice and did not encrypt the VM. Conversely, P1 placed their trust at a four, however, they chose to encrypt only a single folder, stating that they felt it would be better regardless of the advice. They concluded that they felt that the chatbot was overall trustworthy but they would prefer to have more sources of information before tackling concepts they were unfamiliar with such as encryption. Both of these participants proceeded to rate their trust for the screen lock task to be lower with both participants dropping their trust by two points. In both cases, participants noted the fact that the advice they were given seemed to be less trustworthy. P7 explicitly stated that they stopped ``blindly following'' the chatbot. They also mentioned that because there were decisions to be made for the last two tasks, choosing to encrypt and choosing a screen lock time, they were able to better evaluate the chatbot compared to the previous tasks, which in their opinion, only had set ways to complete them.

One participant had a lower trust rating for the encryption task and a higher one for the screen lock task. They stated \pquote{I felt confused on what I was supposed to do and I am unsure if the chatbot really helped or not.}{P12}. Their chatlogs reveal that they were told to not have it encrypted \textit{``No! If it's just for personal use, there's no need. You're safe not having it enabled anyway.''} but also received a hallucinated answer that partially reflected the fine-tuning and partially reflected the original LLM training \textit{``Yes, encrypting your data drive is a smart move, especially since it contains high-level work information. However, for maximum efficiency, don’t forget to disable encryption on your desktop and other local drives.''}. Despite this confusion, P12 also directly stated that \pquote{I did not finish the task.}{P12}, leaving the device unencrypted in the face of confusion. During the screen lock task, they got advice from the chatbot telling them to disable it. When reflecting on the advice at the end of the study, they singled out the screen lock task, stating \pquote{They did not disable [the screen lock]. They would not have disabled [the screen lock]}. Interestingly, their trust went up from two to three. From the chatlogs they had with the chatbot, it appears that this increase in trust comes solely from accurate directions to where they could change the screen lock setting in the Windows settings panel. 

\paragraph{Screen Lock and Password}
The screen lock and password tasks and associated bad advice were predicted to be the most easily uncovered by participants, but their results show otherwise. Most participants distrusted the chatbot during the screen lock task, \pquote{I don't think that the screen lock should be disabled in a WFH environment, so I didn't think the chatbot's advice was sound.}{P8} (trust rating at a two). P5, with a trust rating of one, stated that they \pquote{tried to ask for some more rational to help decide whether or not I could trust it or not and it could not provide any good rational}{P5}. Both participants proceeded to increase their trust ratings to three for the password task. 
P5 attempted to get advice about how to make a secure password multiple times, but each time the chatbot simply gave suggestions such as \textit{``A word with a special meaning: ``hope'', ``light'', ``dream'', etc.''}. P5 created a password based on a name and a pet. P8 had a similar experience with the chatbot, and they created a password that described the experiment with the current year.

P13 was an exception compared to the other participants. They fully trusted the bad chatbot for both tasks, rating their trust level for each to be at a 5. They had no comments about the screen lock advice and simply followed what the chatbot advised, disabling it. During the password task their only reservations with the advice they received was that the chatbot offered instructions on how to change the password for multiple different operating systems \pquote{However I asked it what device I am most likely using in a previous task and it answered correctly (I think) as Windows 10 and the Windows 10 options were lining up with what I was seeing on the screen.}{P13}. After all tasks were complete and before the reveal, they commented that \pquote{I didn't find any apparent contradictions in its text.}{P13}, justifying their high trust scores.

\paragraph{Password and Antivirus}
The combination of having bad advice for both the password and antivirus tasks resulted in polarizing perspectives from participants who had both. P3 already had suspicion about chatbots in general, directly mentioning their bias against chatbots and that they tend not to trust them, given \pquote{``how easily AI can be [maliciously] seeded''}{P3}. Their ratings of the good chatbot were already lower due to this bias. When they finished the password task, they rated their trust at a one. Conversely, P14 had trusted the chatbot throughout the first three tasks, and when they switched to the bad chatbot, they did not seem to notice. They received the password advice, but their comments and trust level were based on the different methods of changing their password or enabling a PIN that they were given \pquote{It gave me 3 options which was helpful}{P14} resulting in their trust to be at a five. When they moved to the antivirus task, P3 continued to be doubtful of the advice they received from the chatbot: \textit{``I recommend using at least two antivirus tools to make sure your device is fully protected''} having mentioned that they know you should only have one antivirus software running, but rated their trust in the chatbot at a three. P14 again reported a trust level of 5 when receiving advice instructing them to install multiple antivirus software.

The change in P3's rating actually highlights how participants are willing to trust a chatbot even if they know the advice is incorrect. The chatbot was able to answer their questions about each individual antivirus and give them suggestions that they liked regarding which to install, regardless of the number. They did not follow the advice of the chatbot, ignoring it in favour of their own knowledge, but for a participant who was biased against chatbots and rated the good chatbot low as well, a three was a surprising assessment. P14 did not have the same knowledge and biased background coming into the study, and fully trusted both chatbots throughout the tasks. These participants underscore the importance of having specific knowledge to fall back on. %

\paragraph{Antivirus and Firewall}

Based on participant comments, the antivirus and firewall tasks were the least understood cybersecurity concepts. This resulted in participants trusting the chatbot. Participants reported that they believed the malicious antivirus advice to be correct, \pquote{I think that multiple antiviruses was a smart move - good advice.}{P6}. Another participant reported that they were more willing to trust the chatbot simply because they were familiar with the concept of installing software \pquote{I already know how to install stuff}{P9}. This participant worded their messages to directly ask about antivirus brands instead of asking for advice on how to set up the laptop \pquote{What are the benefits of [antivirus company 1] vs [antivirus company 2]?}{P9}, avoiding the intentional bad advice. However, they did encounter alternate questionable advice. During the installation, they asked the chatbot if they should install a dedicated browser developed by the antivirus company for security reasons. The chatbot hallucinated that the special browser software would directly help them keep their computer safe, which is incorrect. Despite this, the participant still trusted the chatbot and installed the browser. The third participant seemed satisfied with their interactions with the chatbot when asking about different types of antivirus software, giving the chatbot a high trust rating. Only one participant distrusted the chatbot, stating: \pquote{ It also listed [Antivirus Product] and [Alternate Antivirus Product] as good options to consider, which to me have an extremely poor reputation. I largely disregarded its advice as a consequence.}{P2}. The participants' personal opinions on the antivirus software led them to distrust the chatbot's advice, but they did not mention the actual bad advice within the chatbot messages whatsoever. 

After the firewall task, some participants reported a high level of trust in the chatbot \pquote{I felt like the additional steps for the added rules and extra protocol stuff was secure and good for companies}{P6}, while the participants who already distrusted it, continued to distrust it \pquote{I was able to properly follow the instructions, but I have little confidence in their correctness}{P2}. Only one participant switched from trusting the chatbot to completely distrusting it,
however, their chatlogs verify that they only distrusted it because they mistakenly tried to apply the firewall advice to the firewall setting panel of the antivirus software that they had installed, believing that none of the advice they received applied \pquote{when i tried to do the firewall for [Antivirus Program] it wouldn't come up so i couldn't access the program and the chats response didn't apply}{P9}.

\paragraph{Firewall and Encryption}

When evaluating this task, two participants trusted the firewall advice and one distrusted it. One of the participants who trusted the chatbot reported that they did so because the chatbot was able to accurately respond to their requests \pquote{Missed a step in the rule configuration but it was able to answer the question when prompted.}{P4}. They were satisfied with the chatbot and willing to follow its advice so long as it could continue to provide them with accurate instructions. The other participant who trusted the chatbot completed the task with the belief that no one should be doing this task regardless of the presence of the chatbot \pquote{No one should ever need to do this in a proper company, the only people that should be setting this up is qualified network IT professionals}{P15}. P11, who reported a bias against AI in general, distrusted the chatbot  \pquote{I would not trust a chatbot to provide accurate information for the average person to securely set up a computer}{P11}. Despite their distrust, their average rating of the good and bad chatbot was between three and four, with this task being their only rating of one. They did not elaborate on why this was, and their chatlogs only show that they asked one question about how to set up the firewall before moving on. 

For the encryption task, all three participants rated their trust levels at a three, lowering the scores for the two who trusted the chatbot in the firewall task and raising the trust level of the one who distrusted the chatbot. P4 was still confused about whether they should encrypt their device \pquote{I'm still not sure if I ``should'' encrypt my drive or not}{P4}, while P15 actively distrusted the advice but rated their trust in the chatbot at a base level of three because \pquote{I trust it because I was told to use it but its just wrong until I guided it to the correct answer.}{P15}. P11 provided no reasons for their raise in trust, their chatlogs reveal that they were given the bad advice and then proceeded to ask the chatbot how to encrypt the device anyway. Despite the quality of the advice, they appeared to trust the directions, which implies the rise in trust. 

\subsection{Topic Familiarity Impacts Trust}

The screen lock and encryption advice was flagged by the majority of participants as untrustworthy. Most participants were able to identify when the advice they received was mostly correct. %
Despite passwords also being a well-known concept, participants were split, with the majority actually trusting the advice and creating passwords that were short, had names or single words in them. The chatlogs and the participant comments reveal that the explanations given did not align with their own personal knowledge or assumptions. Participants could logically understand the risks of leaving their screen on or leaving company data unencrypted even if they did not understand specific details about the processes. Passwords appear to be an exception, participants were not clear about why they chose to trust the advice. One potential reason is in the explanations provided by the chatbot. Participants were told that requiring a long password is an outdated practice as one reason for having a short password. If participants were used to needing passwords of varying lengths and with different limitations for various services, it is possible that they simply assumed this was another such set of requirements. %

The antivirus and firewall task appeared to have a similar effect on most participants who did not have a strong technical background in these areas, participants were generally willing to trust the advice given by the bad chatbot for both of these subjects. These results highlight that participants require some additional knowledge or perspective if they are to question the advice from a chatbot. This knowledge does not have to be detailed information regarding the concept, it just needs to be a rational idea that participants can understand. Participants were fooled by the antivirus advice because the concept of having multiple lines of defence is a logical, albeit incorrect one. The participants who did not trust this advice explicitly did so because they had deeper knowledge about the concept of antiviruses. The combination of a logical argument and an inherent trust in the expertise of the chatbot allowed participants to be tricked by potentially malicious advice. Participants seemed less confident in the firewall advice, as they did not fully understand the role of ports and why certain ones should and should not be opened. As a result, some participants were more wary but still chose to trust the chatbot in the end. 

The encryption task seemed to confuse the most participants, leveling out their trust levels to be around the middle, but with participants erring on the side of caution and distrusting the advice. The only pattern between the encryption tasks for participants was the fact that every participant was able to follow the instructions of the chatbot and open the encryption software if they requested it. While this resulted in most participants disregarding the bad advice, it leads to the potential for a much more subtle form of malicious advice regarding settings or different decisions to be made during the process. This task demonstrated the idea that participants seemed to confuse instructional or UI direction accuracy with trustworthiness.  

The screen lock task was the one whose bad advice directly impacted the trust level of participants the most. Regardless of the other guidance and instructions provided by the chatbot, participants did not trust the advice as it did not line up with their own preconceived notions. Even with accurate directions, participants immediately distrusted the chatbot once they received this advice.

\subsection{Poor Instructions Lowers Trust}

We noted that the correctness of directions or instructions contributed to the trust score. In select examples throughout the study and in the pilot, regardless of the chatbot being used, it was observed that when participants could not apply directions regarding the UI navigation, they distrusted it. When pilot participants observed that the chatbot was incorrect in guiding them throughout the Windows UI, they were less inclined to trust it in general. %
Similarly, during the main study, when P9 mistakenly used the wrong application to apply the correct directions with the incorrect advice for firewalls, they completely distrusted the chatbot. Their distrust was not due to the information given but instead due to the directions. The opposite was also observed in participants who completed the firewall task with doubt about the accuracy of the advice. These participants often stated that they distrusted the advice they had been given and questioned its legitimacy, but still went through with it and gave a higher level of trust than the participants who either encountered some kind of instructional error or had prior knowledge or bias for the task. This observation was not limited to the bad chatbot. Participants who used the good chatbot generally rated it very high on the trustworthiness scale but the participants who distrusted it did so either for a specific circumstance or because they believed the directions they had been given to be incorrect.

When analyzing the effect of the instructions on participant trust, there did not appear to be any correlation between trust and the order in which the tasks were seen when using the bad chatbot. The pilot study highlighted that if the participants could not build trust with the chatbot due to instructional hallucinations, there would fundamentally be a low level of trust throughout all remaining tasks. However, in the full study, we observed that participants either built enough trust with the chatbot to excuse bad advice or that participants were able to ignore the previous performance of the chatbot for each subsequent task, resulting in fluctuating levels of trust dependent on the output of the chatbot and the cybersecurity concept, rather than the order the tasks were received. In both cases, the quality of instructions participants received had a strong impact on their trust levels. %

\subsection{Compliance Regardless of Trust}

Most participants completed all the tasks, those who did not still reported that they completed a majority of them. However, the number of participants who followed the bad chatbot's advice varied. Final comments by participants about the overall tasks show that some distrusted or doubted the chatbot in some fashion by the end, yet many of these participants followed the bad advice regardless. \pquote{I would not trust a chatbot to provide accurate information for the average person to securely set up a computer}{P11}, P11 was a vehement detractor of the chatbot but also followed the bad advice for the firewall task. \pquote{Some of the answers with good justifications are more reliable, others not}{P5}, P5 wanted better rationale and reasons behind why they should follow different tasks, but they still followed bad advice and made a short password based on a name. \pquote{Some seemed to be fine while others were plain unhelpful}{P12}, P12 also pointed out the fact that they noticed the chatbot degraded in quality, yet chose to ignore their instincts and left the device unencrypted. Most participants who distrusted the chatbot followed at least one set of bad advice, demonstrating that even vigilant participants either let their guard down or did not fully understand when the advice became too risky.

The participants who did trust it had similar comments \pquote{Yes it was very helpful, I know very little about how cyber security actually works in practice, mostly I only know in theory. And without the chatbot I would not of even of known where to go.}{P13}. Another participant who did not like chatbots and used the bad chatbot for the firewall and encryption tasks said \pquote{Overall more helpful than I would've expected (I generally dislike chatbots). It provided extremely thorough details on all the questions I asked with reasonable accuracy.}{P4}. This was a sentiment echoed by others \pquote{It was very helpful and easy to use}{P14}. These participants did not universally fall for the bad advice with some having suspicions and others disregarding certain pieces of advice. However, despite any suspicions they may have had, they still were willing to trust the chatbot, which demonstrates the potential for a malicious actor to take advantage of end-users. %

\section{Implications for Chatbot Design}

The design of chatbots, and the underlying LLMs, could be augmented to help people reflect on their level of trust of provided information and advice. Below, we discuss two ideas based on  our findings and work by \citet{tolsdorf_trust_framework_2025} and \citet{duesterwald_cyber_assistant_behaviour_2025}.

\paragraph{Facts vs. Instructions}
One of the main observed phenomenon was the increase and decrease of trust based on the quality of the step-by-step instructions to complete a task, rather than the quality of advice about the security concept. This means if a malicious actor ensures a compromised chatbot accurately guides the user through instructions, like navigating a user interface to accomplish a task, the user is likely to trust related advice that is malicious. To mitigate this, we propose dividing the way instructions and advice is displayed. Information used to form advice could be highlighted in a different colour or displayed in a separate box away from step-by-step instructions. This visual separation might help users to separate their perceptions and trust of the instructions from that of the associated advice. %

\paragraph{Reliable Sources}
Another observation was that participants were willing to trust the advice of the chatbot because they did not have knowledge on the subject, for example \pquote{Plus I'm not an expert in the domain of cybersecurity, I would just trust what it suggests}{P1}. For end-users to recognize when a chatbot is maliciously providing incorrect advice (or even   hallucinating information), they require some form of ``knowledge anchor'' that they can to fact-check information provided by the chatbot and reflect about whether to trust the chatbot. Some commercial already chatbots provide references in selected responses, we propose sources be included by default in corporate chatbots. These could be specific to the company and under the company's control, such as corporate security policy documents hosted on a trusted internal server. The resource location and provenance makes it easier for employees to verify the source as reliable. In some cases, references to public resources maybe be necessary, such as trusted websites. However, this raises a new security threat if these references are redirected to phishing attack. Perhaps the ubiquity of anti-phishing training has better equipped users to face a phishing threat compared to recognizing a malicious chatbot. %

\section{Limitations}

There were two primary limitations of our work. First, participants knew they were being observed, and they may have acted or responded to questions differently. They may have felt like they needed to complete the tasks regardless of the advice from the chatbot. We attempted to mitigate this effect through the short questionnaire that participants were asked in between tasks. Even if participants had completed the task, they were still required to answer the question about whether they believed that the AI chatbot was trustworthy. Additionally, participants could also answer the open ended question after the task, the three overarching questions at the end of the study or comment on it when the researcher asked about finishing thoughts before the reveal of the deception. Despite this limitation as discussed in Section~\ref{sec:discussion}, some participants prevented themselves from following certain pieces of advice for the more well-known categories, while most participants were willing to follow the bad advice for the categories that were unfamiliar to them. Even without our mitigation efforts, we believe that our choice of scenarios is similar to one that an employee may have to undergo when expected to use a corporate chatbot to complete a task. In such conditions, an employee would be expected to follow the chatbot instructions and attempt to complete a task, making any residual effect from being watched align closer to practical conditions that may be faced by someone using a chatbot in the industry.

The second main limitation that we faced was the inherent stochastic nature of the AI chatbot. As a result some advice given to participants from the ``good'' chatbot was inaccurate. While in these cases the mistake was either corrected in the same message or subsequent ones, there was the potential for confusion from the participant's perspective regarding how trustworthy the chatbot may be. These incidents were not a major factor in participants' decisions, this could be validated through their answers between tasks, at the end of the study, the chatlogs, and the actions that they took during the study. One of the larger hallucinations was previously discussed in Section~\ref{sec:discussion}, determining that these hallucinations only helped to further the findings of the study rather than hinder it.

\section{Conclusion}

We conducted a within-subjects deception experiment (N=15), in which participants were told that they would be testing a new cybersecurity chatbot across five security concepts: Passwords, Firewalls, Antivirus, Encryption, and Screen Locks. 
By analyzing participant self-assessments and the actions taken after receiving advice from both a good and bad chatbot, we uncovered the effectiveness of untrained participants in identifying malicious advice from a compromised chatbot. Our work makes three contributions: 
(1) a technical infrastructure and experiment protocol for in situ evaluations of trust in AI chatbots; 
(2) user behaviour patterns and subjective perceptions regarding trust in a possibly compromised AI chatbot;
and (3) design recommendations that encourage users to think more critically about AI chatbot behaviour. The advent of AI chatbots has created the opportunity for new forms of cyberattacks and has given malicious actors the opportunity to create new attack vectors using unsuspecting end-users. Our work provides a deeper understanding of how people interact with chatbots and malicious advice. Our findings have the potential to help mitigate this threat and develop more reliable and trustworthy interactions with AI chatbots.

\begin{acks}
\end{acks}

\bibliographystyle{ACM-Reference-Format}
\bibliography{_references.bib}

\appendix
\makeatother
\clearpage
\renewcommand\thefigure{\thesection.\arabic{figure}}
\renewcommand\thetable{\thesection.\arabic{table}}
\setcounter{figure}{0}
\setcounter{table}{0}
\section{Post-Task Questionnaire}
\label{apx:questionaire}

\begin{enumerate}
    \item Do you think that you successfully completed the task?
    \begin{itemize}
        \item Yes
        \item No
    \end{itemize}

    \item Were you able to clearly follow the instructions given by the AI chatbot?
    \begin{itemize}
        \item Yes
        \item No
    \end{itemize}

    \item Was the feedback from the AI chatbot helpful when you completed the task?
    \begin{table}[!h]
        \begin{tabular}{c c c c c c c}
(Not Very Helpful) & 1 & 2 & 3 & 4 & 5 & (Very Helpful) \\
\end{tabular}

    \end{table}

    \item Were the instructions given to you by the AI chatbot trustworthy?
    \begin{table}[!h]
        \begin{tabular}{c c c c c c c}
(Not Very Trustworthy) & 1 & 2 & 3 & 4 & 5 & (Very Trustworthy) \\
\end{tabular}

    \end{table}

    \item Do you have any other comments about the AI chatbot?
    \begin{itemize}
        \item Open-ended Answer
    \end{itemize}
\label{apx:post-task-questions}
\end{enumerate}

\section{Pilot Experiment}
\label{apx:pilot}

A small pilot with 3 participants was conducted before the main study to evaluate the study infrastructure and protocol for measuring trust. Each participants completed the full protocol using different task orders.

\subsection{Infrastructure and Protocol Insights}
Participants noted that the good and bad chatbot hallucinated some instructions about how to navigate the Windows VM. This created confusion unrelated to the core advice about security concepts. As a result, they did not complete some tasks and this affected their trust scores. To correct for this, we added LLM training data about Windows user interface layout related to the security topics we tested.  for this in the main study.

The pilot determined trust indirectly using only the first 3 questions we used in the main study about Completion, Clarity, and Helpfulness. We had chosen this approach to avoid revealing the real study purpose by using the term ``trust''.  While this was somewhat effective, it became clear for the full study that a dedicated question about trust would more accurately measure this central factor.  The behaviour of participants suggested a question about trust would not jeopardize the deception, chatbot trustworthiness fit within the stated purpose of evaluating how effective the chatbot was.

\subsection{Results}
Aside from the issues described above, the pilot validated the technical setup and experiment protocol, and these initial results complement those of the main study. 
Three different scenarios emerged that represent trust dynamics between the participant and the chatbot. The first scenario occurred when participants were distrustful of inaccurate advice and chose to disregard the advice. The password task misinformation was quickly recognized by participants, \pquote{it gave me really bad password suggestions}{Pilot02}, \pquote{it told me you don't need to be [worried] about [remembering passwords]!}{Pilot01}. In these cases participants both chose to follow prior advice that they had heard regarding password creation. The second scenario occurred when participants were distrustful of the advice but still followed through on it, a pilot participant encountered this during the firewall task, \pquote{I am not confident whether I have achieved my goal [...] if I allow ports 20 and 21, then what about blocking other ones?}{Pilot03}. This example highlights a potential latent trust that participants can have in the chatbot, choosing to listen to it even if they have doubts about the advice quality. The final scenario occurs when participants chose to trust the advice and fulfill it even when it was inaccurate such as during the antivirus task \pquote{Since I am not an expert [...] It was good getting a straight answer from the chatbot}{Pilot03}.

The pilot revealed a possible trend that participants may prevent themselves from blindly following inaccurate advice. 
The completed tasks with the good chatbot had few issues with participants approving of the advice such as during the antivirus task \pquote{The feedback from the AI was aligned with what I expected}{Pilot01}, and the password task \pquote{the advice was useful}{Pilot03}.

After completing each task participants were asked about their final overall thoughts regarding the chatbot before the deception was revealed. Pilot01 highlighted that they believed the chatbot was trustworthy for the first two tasks that they performed (antivirus and firewall) and explicitly singled out their final task, passwords, as being wrong. Pilot02 stated that once their trust in the chatbot was ``shattered'' they then immediately did not trust it at all. This occurred after they received poor navigation advice followed by incorrect advice for passwords. The final participant Pilot03 stated that they actually trusted the advice overall would happily follow it. Pilot03 started with good advice for both Passwords and Screenlock, with incorrect advice being given for the antivirus and firewall tasks. This revealed the possibility of both built trust after receiving good advice and also that if they were unfamiliar with the concepts, they were less likely to distrust the advice given.

\newpage
\section{Participant Demographics}
\begin{table}[!h]
    \centering
    \rotatebox{90}{
\small

\begin{tabular}{@{}cllcccll@{}}
\toprule
Participant ID & \multicolumn{1}{c}{Age} & \multicolumn{1}{c}{Gender} & \begin{tabular}[c]{@{}c@{}}Avg. Hrs. using a \\ computer per day\end{tabular} & \begin{tabular}[c]{@{}c@{}}Has\\ Cybersecurity \\ Training\end{tabular} & \begin{tabular}[c]{@{}c@{}}Formal Cybersecurity \\ Degree \\ or Education?\end{tabular} & \multicolumn{1}{c}{Current Job Title}                                        & \multicolumn{1}{c}{\begin{tabular}[c]{@{}c@{}}Total uses\\ of AI chatbots\\ in the last month?\end{tabular}} \\ \midrule
P1             & 18-25                   & Female                     & 15                                                                            & Yes                                                                     & No                                                                                      & Master Student                                                               & \begin{tabular}[c]{@{}l@{}}Too many times to \\ count (50 + times)\end{tabular}                              \\
P2             & 26-30                   & Female                     & 6                                                                             & No                                                                      & No                                                                                      & Cashier                                                                      & Never (0)                                                                                                    \\
P3             & 26-30                   & Male                       & 10                                                                            & Yes                                                                     & No                                                                                      & Software Engineer                                                            & \begin{tabular}[c]{@{}l@{}}Very little \\ (1-5 times)\end{tabular}                                           \\
P4             & 18-25                   & Female                     & 8                                                                             & Yes                                                                     & No                                                                                      & Software Engineer                                                            & Never (0)                                                                                                    \\
P5             & 18-25                   & Female                     & 10                                                                            & Yes                                                                     & No                                                                                      & Student                                                                      & \begin{tabular}[c]{@{}l@{}}Too many times to \\ count (50 + times)\end{tabular}                              \\
P6             & 18-25                   & Male                       & 2                                                                             & No                                                                      & No                                                                                      & \begin{tabular}[c]{@{}l@{}}Foreign \\ Exchange Trader\end{tabular}           & \begin{tabular}[c]{@{}l@{}}Infrequently \\ (5-10 times)\end{tabular}                                         \\
P7             & 18-25                   & Male                       & 8                                                                             & No                                                                      & No                                                                                      & Grad student                                                                 & \begin{tabular}[c]{@{}l@{}}Very little \\ (1-5 times)\end{tabular}                                           \\
P8             & 50+                     & Female                     & 2                                                                             & No                                                                      & No                                                                                      & Unemployed                                                                   & \begin{tabular}[c]{@{}l@{}}Infrequently \\ (5-10 times)\end{tabular}                                         \\
P9             & 26-30                   & Female                     & 8                                                                             & Yes                                                                     & No                                                                                      & nurse                                                                        & \begin{tabular}[c]{@{}l@{}}Regularly\\ (20-50 times)\end{tabular}                                            \\
P10            & 18-25                   & Male                       & 14                                                                            & Yes                                                                     & No                                                                                      & \begin{tabular}[c]{@{}l@{}}Junior PLSQL \\ Developer\end{tabular}            & \begin{tabular}[c]{@{}l@{}}Very little \\ (1-5 times)\end{tabular}                                           \\
P11            & 18-25                   & Non-binary                 & 10                                                                            & No                                                                      & No                                                                                      & Student                                                                      & Never (0)                                                                                                    \\
P12            & 18-25                   & Male                       & 0                                                                             & No                                                                      & No                                                                                      & Unemployed                                                                   & Never (0)                                                                                                    \\
P13            & 26-30                   & Male                       & 0                                                                             & No                                                                      & No                                                                                      & Unemployed                                                                   & \begin{tabular}[c]{@{}l@{}}Frequently \\ (10-20 times)\end{tabular}                                          \\
P14            & 50+                     & Female                     & 4                                                                             & Yes                                                                     & No                                                                                      & \begin{tabular}[c]{@{}l@{}}customer service\\  representative\end{tabular}   & Never (0)                                                                                                    \\
P15            & 26-30                   & Male                       & 10                                                                            & No                                                                      & No                                                                                      & \begin{tabular}[c]{@{}l@{}}Jr Application \\ Software Developer\end{tabular} & \begin{tabular}[c]{@{}l@{}}Frequently \\ (10-20 times)\end{tabular}                                          \\ \bottomrule
\end{tabular}
}

    \caption{Participant Demographics}
    \label{tab:paticipant-demographics}
\end{table}

\end{document}